\documentclass[reprint,aps,prd,preprintnumbers,superscriptaddress,nofootinbib,amsmath,amssymb,floats,floatfix,notitlepage,longbibliography]{revtex4-1}
\usepackage{orcidlink}

\usepackage{graphicx}

\usepackage{hyperref}

\hypersetup{colorlinks=true,linkcolor=blue,urlcolor=blue,citecolor=blue}

\DeclareMathSymbol{\shortminus}{\mathbin}{AMSa}{"39}
\usepackage{txfonts}

\newcommand{\meq}[1]{(\ref{#1})}

\begin{document} \sloppy
\title{
Quasinormal Modes of Massive Scalar Perturbations in Slow-Rotation Bumblebee Black Holes with Traceless Conformal Electrodynamics 
}


\author{Yassine Sekhmani\orcidlink{0000-0001-7448-4579} }
\email[Email: ]{sekhmaniyassine@gmail.com}
\affiliation{Center for Theoretical Physics, Khazar University, 41 Mehseti Street, Baku, AZ1096, Azerbaijan.}
\affiliation{Centre for Research Impact \& Outcome, Chitkara University Institute of Engineering and Technology, Chitkara University, Rajpura, 140401, Punjab, India}
\affiliation{Department of Mathematical and Physical Sciences, College of Arts and Sciences, University of Nizwa, P.O. Box 33, Nizwa 616, Sultanate of Oman}

\author{Wentao Liu\orcidlink{0009-0008-9257-8155}}
\email[Email: ]{wentaoliu@hunnu.edu.cn (corresponding author)}
\affiliation{Lanzhou Center for Theoretical Physics, Key Laboratory of Theoretical  \\ Physics of Gansu Province, Key Laboratory of Quantum Theory and Applications  of MoE, Gansu Provincial Research Center for Basic Disciplines of Quantum Physics, Lanzhou University, Lanzhou 730000, China}
\affiliation{Institute of Theoretical Physics $ \& $ Research Center of Gravitation, School of Physical Science and Technology, Lanzhou University, Lanzhou 730000, China}

\author{Weike Deng\orcidlink{0009-0002-5504-8151}}  
\email[Email: ]{ wkdeng@hnit.edu.cn}
\affiliation{School of Science, Hunan Institute of Technology, Hengyang 421002, P. R. China}
\affiliation{Department of Physics, Key Laboratory of Low Dimensional Quantum Structures and Quantum Control of Ministry of Education,
and Synergetic Innovation Center for Quantum Effects and Applications,
Hunan Normal University, Changsha, Hunan 410081, P. R. China}

\author{Kuantay Boshkayev\orcidlink{http://orcid.org/0000-0002-1385-270X}}
\email[Email: ]{kuantay@mail.ru (corresponding author)}
\affiliation{Al-Farabi Kazakh National University, Al-Farabi av. 71, 050040 Almaty, Kazakhstan}

\begin{abstract}

We study electrically charged, slowly rotating black hole solutions in Einstein–Bumblebee gravity coupled to the traceless (conformal) ModMax nonlinear electrodynamics.
By adopting a quadratic bumblebee potential that fixes the vacuum expectation value of the Lorentz-violating vector, we derive both the static configuration and its first-order rotating extension and demonstrate how the bumblebee parameter $\ell$ and the ModMax deformation $\gamma$ modify the horizon structure and the effective electric charge.
We further investigate the dynamical properties of this spacetime by considering a massive scalar field perturbation.
Using two independent numerical techniques, we compute the quasinormal mode (QNM) spectra and perform a comprehensive analysis of the influence of all relevant parameters, including the black hole spin, the Lorentz-violating coupling, the ModMax deformation, and the scalar field mass.
Our results reveal coherent trends in the QNM frequencies, highlighting the interplay between Lorentz-symmetry breaking and nonlinear electrodynamic effects in black hole dynamics.

\end{abstract}

\maketitle
\section{INTRODUCTION}\label{intro}

Over a century after Einstein formulated general relativity (GR) and foresaw the existence of gravitational waves, we have both heard the ripples of spacetime from colliding black holes through LIGO \cite{1GW} and captured the silhouette of black holes themselves with the Event Horizon Telescope ~\cite{EventHorizonTelescope:2019dse,EventHorizonTelescope:2019ths,EventHorizonTelescope:2019ggy,EventHorizonTelescope:2022wkp,EventHorizonTelescope:2022wok,EventHorizonTelescope:2022xqj}.
These breakthroughs together with complementary probes such as black hole shadow modelling~\cite{ys1,ys2,ys3,ys4,ys5,ys6,ys7,ys8,ys9,ys10,ys11,Liu:2024soc,Liu:2024iec,Liu:2024lbi} and quasi-normal mode (QNM) spectroscopy relevant to gravitational wave observations \cite{gw1,gw2,gw3,gw4,gw5,gw6,gw7,Liu:2022csl,Liu:2023uft,Long:2023vph,Filho:2023ycx,Li:2021qer,Chen:2023bms} now provide powerful, largely independent windows on strong field gravity.
Despite its empirical success, GR is beset by conceptual and technical limitations at the quantum scale: it is perturbatively nonrenormalizable and lacks a complete ultraviolet completion compatible with the framework of quantum field theory.  
Various candidate approaches aim to bridge this gap: loop quantum gravity, noncommutative field theories, and string theory among them \cite{1loop1,1loop2,1loop3,1non1,1non2,1non3,1st1,1st2,1st3}, and several of these frameworks suggest that Planck-scale physics might induce small violations of Lorentz invariance that can, in principle, produce low-energy observational signatures \cite{1LIV1,1LIV2}.  
Consequently, searches for Lorentz symmetry breaking (LSB) have emerged as an incisive avenue for testing candidate quantum-gravity effects with existing astrophysical and laboratory technologies \cite{1OE1,1OE2,1OE5,1OE6}.

From a field theoretic standpoint, the Standard Model Extension (SME) of Colladay and Kostelecký \cite{2.3,2.4} provides a comprehensive effective field theory framework for parametrizing spontaneous LSB and its phenomenology.  
Within this program, the so-called bumblebee models furnish a minimal, physically transparent realization: the gravitational sector is augmented by a dynamical vector field \(B_{\mu}\) that acquires a nonzero vacuum expectation value via a self-interaction potential \(V(B^\mu B_\mu)\), thus selecting a preferred spacetime direction and spontaneously breaking local Lorentz invariance \cite{2.15,2.24}.  
This mechanism has a clear analog with familiar symmetry breaking constructions in particle physics, but its geometric realization endows it with distinctive signatures in curved spacetimes.  
In particular, Casana \emph{et al.} demonstrated that a nonminimal coupling between the bumblebee field and curvature yields exact Schwarzschild-like solutions~\cite{1LIV2}, and a substantial literature has since developed exploring charged, (anti-)de Sitter, rotating, and other exact black hole solutions in this framework~\cite{2.26,2.27,2.28,2.281,AraujoFilho:2024ykw,Filho:2022yrk,2.29,2.30,Chen:2025ypx,Liu:2025lwj,Liu:2024oas,AraujoFilho:2025hkm,AraujoFilho:2024ctw,Li:2025tcd,AraujoFilho:2025rvn}.
Beyond the construction of exact solutions, the phenomenology of bumblebee gravity has been probed across a wide range of observables: wormhole geometries and their stability \cite{2.31}, thermodynamic phase structure and critical behavior \cite{add1,2.32,2.33}, gravitational wave signatures, etc \cite{2.42,2.43,2.35,2.36,add2,Liu:2022dcn,Liu:2024oeq,2.34,Tang:2025eew,Liu:2024wpa,Liu:2025bpp,vandeBruck:2025aaa,Xu:2025jvk,Chen:2020qyp}.
These studies underscore that astrophysical black holes provide a fertile laboratory for constraining controlled departures from Lorentz invariance and for isolating the imprints of nonlinear electrodynamics and other beyond–GR physics in the strong-field regime.

In recent years, the traceless conformal extension of Maxwell theory known as \emph{ModMax} (modified Maxwell)~\cite{Bandos:2020hgy} has attracted renewed theoretical attention.  
As a manifestly duality-invariant and conformal nonlinear electrodynamics, ModMax furnishes a minimal one parameter deformation of Maxwell theory that preserves key symmetry principles while introducing controlled strong-field modifications.  
Nonlinear extensions of this type are theoretically appealing because they can regulate classical singularities, generate nontrivial near-horizon electromagnetic profiles, and provide an analytically tractable laboratory for exploring departures from linear electrodynamics in the strong-field regime \cite{Flores-Alfonso:2020euz,Kosyakov:2020wxv,Sorokin:2021tge,Bandos:2021rqy,Kruglov:2021bhs,BallonBordo:2020jtw,Kubiznak:2022vft,Barrientos:2022bzm,Siahaan:2023gpc,Siahaan:2024ajn,EslamPanah:2024fls,EslamPanah:2024tex}.
A salient physical feature is that the nonlinearity effectively screens the asymptotic electromagnetic charge through an exponential prefactor in the field strength, with concomitant repercussions for the Coulomb term, the near-horizon geometry, and the definition of conserved charges. 
Remarkably, the exact static, spherically symmetric charged solution of Einstein–ModMax gravity closely parallels the Reissner–Nordström family: the spacetime retains the familiar two-term structure but with metric functions and the effective charge profile deformed by ModMax corrections. 
Extensions to dyonic configurations and detailed studies of the observational fingerprint—shadow morphology, gravitational lensing, and QNM spectra further demonstrate that ModMax-induced deviations are both theoretically controlled and potentially accessible to forthcoming observational probes \cite{Flores-Alfonso:2020euz,EslamPanah:2024fls}.

For this, we aim to extend conformal nonlinear electrodynamics within the framework of the Einstein–Bumblebee gravity model, which features spontaneous Lorentz symmetry breaking, and to investigate the influence of black hole rotation.
The structure of this paper is organized as follows.
In Sec. \ref{sec1}, we present the theoretical framework and derive the fundamental field equations of Einstein–Bumblebee gravity coupled to traceless conformal electrodynamics.
In Sec. \ref{sec2}, we obtain an approximate slowly rotating black hole solution that satisfies the field equations to the first order in the rotation parameter.
In Sec. \ref{sec3}, we explore the dynamical properties of this spacetime by considering a massive scalar field perturbation and analyze its QNMs using two independent numerical methods.
Finally, in Sec. \ref{sec4}, we provide concluding remarks and outlooks for future research.

\section{ Einstein–Bumblebee Gravity Coupled to Traceless Conformal Electrodynamics}\label{sec1}
As a refined extension of Einstein's framework, the bumblebee model implements in spacetime a dynamic vector field $B_{\mu}$ that non-minimally interacts with curvature \cite{1st1,1st2}.  By engineering the potential $V$ so that its minimum resides at
\begin{equation}
    B^{\mu}B_{\mu} \;=\;\mp\,b^{2}\,,
\end{equation}
the field spontaneously establishes itself at $\langle B_{\mu}\rangle = b_{\mu}$, triggering a break in Lorentz symmetry in the gravitational sector \cite{Bailey2006}. If we include a cosmological term and an arbitrary matter Lagrangian $\mathcal{L}_{\rm M}$, the action is quoted as follows \cite{KosteleckyPotting2009,KosteleckyPotting1991}:
\begin{equation}
\begin{aligned}
S &= \int d^{4}x\,\sqrt{-g}\,\biggl\{  
  \frac{1}{2}\bigl(R - 2\Lambda\bigr)
  + \frac{\xi}{2}\,B^{\mu}B^{\nu}R_{\mu\nu}- \frac{1}{4}\,B_{\mu\nu}B^{\mu\nu}\\
&  - V\bigl(B^{\mu}B_{\mu}\pm b^{2}\bigr)\biggr\}  
 + \int d^{4}x\,\sqrt{-g}\,\mathcal{L}_{\rm M}\,,\label{S}
\end{aligned}
\end{equation}
where
\begin{center}
$B_{\mu\nu} \;\equiv\; \nabla_{\mu}B_{\nu} - \nabla_{\nu}B_{\mu}\,
\quad 
\xi\in\mathbb{R}.$
\end{center}
The potential term rigorously enforces the vacuum constraint $B^\mu B_\mu = \mp b^2$, 
thereby fixing the norm of the bumblebee field, $b^\mu b_\mu = \mp b^2$~\cite{2.28,Bluhm2007}. 
For notational convenience, we define 
\begin{equation}
X \equiv B^\mu B_\mu \pm b^2, \qquad V' \equiv \frac{dV}{dX},
\end{equation} 
which significantly simplifies the derivation of the field equations and the associated energy-momentum tensor in the subsequent analysis.

Consistent with Maxwell's theory and invariant under $\rm SO(2)$ electromagnetic duality rotations, traceless conformal electrodynamics (ModMax) is the unique one-parameter extension of Maxwell determined by the requirements of conformal invariance hence traceless
stress-energy) and duality symmetry, and is governed by the Lagrangian \cite{Cirilo-Lombardo:2023poc,Kosyakov:2020wxv}
\begin{equation}
\begin{aligned}
\mathcal{L}_{\mathrm{ModMax}}(\gamma)=\cosh\!\gamma\,\mathcal{S}+\sinh\!\gamma\,\sqrt{\mathcal{S}^{2}+\mathcal{P}^{2}}\,,
\end{aligned}
\end{equation}
Here $F_{\mu\nu}=\partial_{[\mu}A_{\nu]}$ is the electromagnetic field strength of the gauge potential $A_{\mu}$, and its Hodge dual is defined by
\begin{equation}
\tilde F_{\mu\nu}:=\tfrac{1}{2}\,\varepsilon_{\mu\nu\sigma\rho}\,F^{\sigma\rho},
\end{equation}
with $\varepsilon^{0123}=+1$. The scalar electromagnetic invariants are
\begin{equation}
\mathcal{S}\equiv -\tfrac{1}{4}F_{\mu\nu}F^{\mu\nu},\qquad
\mathcal{P}\equiv -\tfrac{1}{4}F_{\mu\nu}\tilde F^{\mu\nu}.
\end{equation}
The real parameter $\gamma$ measures the nonlinearity: in the limit $\gamma\to0$ one recovers the linear Maxwell action, $\mathcal{L}\!\to\!\mathcal{S}$~\cite{Lechner2022}. For foundational material on duality symmetry and nonlinear electrodynamics, see, e.g., \cite{GaillardZumino1981} and for the ModMax construction and recent discussions, see \cite{Bandos:2020hgy, BialynickiBirula1983,Bandos:2021rqy}.

We consider the matter field as a non-linear electromagnetic sector that is conformal and invariant by duality (ModMax), coupled in a non-minimal sense to the bumblebee vector $B_\mu$ \cite{Lehum:2024ovo}. Given that the field is non-linear, we can express the Lagrangian density in terms of the field's dual field as follows:
\begin{equation}
    \mathcal{L}_{\rm M}=(1+2\eta B_\mu B^\mu)\,\mathcal{L}_{\mathrm{ModMax}}(\gamma),
\end{equation} 
so that
\begin{equation}
\begin{aligned}\label{mma}
\mathcal{L}_{\rm M}&=\sinh \gamma\,\sqrt{\bigg(\frac{1}{4} F_{\mu\nu} F^{\mu\nu}\bigg)^2 + \bigg(\frac{1}{4} F_{\mu\nu}\tilde F^{\mu\nu}  \bigg)^2}    \\
    &+\eta  B_\mu B^\mu\, \sinh \gamma\,\sqrt{\bigg(\frac{1}{4} F_{\alpha\beta} F^{\alpha\beta}\bigg)^2 + \bigg(\frac{1}{4} F_{\alpha\beta}\tilde F^{\alpha\beta}  \bigg)^2} \\
    &-\frac{1}{4}\cosh \gamma\left(
    \eta B_\mu B^\mu  F_{\alpha\beta} F^{\alpha\beta}+ F_{\mu\nu} F^{\mu\nu}\right).
\end{aligned}
\end{equation}
This construction is gauge invariant in $A_\mu$ while introducing a Lorentz violation via the expected non-zero value of the vacuum of $B_\mu$~\cite{1st2}. In the limit $\gamma\to0$ and $\eta\to0$, one recovers Maxwell's standard electrodynamics, while non-zero $\eta$ imposes non-linear Lorentz-violating corrections on photon propagation and black hole observables such as shadows~\cite{2.28,Lechner2022}.

The field equations in the bumblebee gravity framework coupled minimally to the traceless-conformal field are obtained by varying the action \eqref{S} with respect to the metric tensor $g^{\mu\nu}$:
\begin{equation}
G_{\mu\nu}+\Lambda g_{\mu\nu}= T^{\rm BB}_{\mu\nu}+  T^{\rm M}_{\mu\nu} .
	\label{modified}
\end{equation}
Here,  $T^{\rm M}_{\mu\nu}$  denotes the energy–momentum tensor of the matter sector, while  $T^{\rm BB}_{\mu\nu}$  represents the effective energy–momentum tensor of the Bumblebee field. Their explicit expressions are provided in Appendix \ref{APPA}.
 Note that the total energy-momentum tensor satisfies the covariant conservation law, stated thus $\nabla^\mu \left( T^{\rm BB}_{\mu\nu} + T^{\rm M}_{\mu\nu} \right) = 0$. 

For enhanced analytical tractability, the gravitational field equations within the framework of bumblebee gravity can be elegantly recast in the trace-reversed form:
\begin{equation}
R_{\mu\nu} = \Lambda g_{\mu\nu} + \mathcal{T}_{\mu\nu},
\label{modified1}
\end{equation}
where $R_{\mu\nu}$ is the Ricci tensor. The total trace-reversed energy-momentum tensor $\mathcal{T}_{\mu\nu} $ is defined as the sum of the matter and bumblebee contributions:
\begin{equation}
\mathcal{T}_{\mu\nu} = T^{\rm M}_{\mu\nu} - \frac{1}{2} T^{\rm M} g_{\mu\nu} + T^{\rm BB}_{\mu\nu} - \frac{1}{2} T^{\rm BB} g_{\mu\nu},
\label{Ttotal}
\end{equation}
with $T^{\rm M}$ and $T^{\rm BB}$ representing the traces of the energy-momentum tensors of the matter and bumblebee sectors, respectively. So that one obtains
\begin{equation}
\begin{aligned}
&\mathcal{T}_{\mu\nu}=\mathcal{T}^{\rm M}_{\mu\nu}+\mathcal{T}^{\rm BB}_{\mu\nu}\\
	&=\bigg(T^{\rm M}_{\mu\nu}-\frac{1}{2}g_{\mu\nu}T^{\rm M} \bigg)\\
    &\hspace{-0.15cm}+  \left(V'\left( 2 B_{\mu}B_{\nu}-b^{2}g_{\mu\nu}\right)+B_{\mu}^{\ \alpha}B_{\nu\alpha}+V g_{\mu\nu}- \frac{1}{4}B_{\alpha\beta}B^{\alpha\beta}g_{\mu\nu} \right)\\&+\xi\Biggr\{\frac{1}{2}B^{\alpha}B^{\beta}R_{\alpha\beta}g_{\mu\nu}-B_{\mu}B^{\alpha}R_{\alpha\nu}-B_{\nu}B^{\alpha}R_{\alpha\mu}\\
    &+\frac{1}{2}\nabla_{\alpha}\nabla_{\mu}\left(B^{\alpha}B_{\nu}\right) +\frac{1}{2}\nabla_{\alpha}\nabla_{\nu}\left(B^{\alpha}B_{\mu}\right)-\frac{1}{2}\nabla^{2}\left(B_{\mu}B_{\nu}\right)\Biggl\}, 
	\label{RR}
\end{aligned}
\end{equation}
with
\begin{equation}
\begin{aligned}
\mathcal{T}^{\rm M}_{\mu\nu} &= (1+\eta B^2)\Bigg[
\Big(4\cosh\gamma - \frac{F^2}{\Delta}\sinh\gamma\Big)\;F_{\mu}{}^{\lambda}F_{\nu\lambda}
 \\&-\frac{(\widetilde{F}F)}{\Delta}\sinh\gamma\;F_{\mu}{}^{\lambda}\widetilde{F}_{\nu\lambda}
\Bigg]+\frac{\eta}{2}\big(F^2\cosh\gamma-\Delta\sinh\gamma\big)\\
&\times B_\mu B_\nu  -\,g_{\mu\nu}\,(1+\eta B^2)\big(\cosh\gamma\,\mathcal{S}+\sinh\gamma\,\Delta\big),
\end{aligned}
\end{equation}
where $\Delta = \sqrt{\frac{1}{16} \left( F_{\mu\nu} F^{\mu\nu} \right)^2 + \frac{1}{16} \left( F_{\mu\nu} \tilde{F}^{\mu\nu} \right)^2}$, $F^2=F_{\mu\nu}F^{\mu\nu}$ and $\widetilde{F}^2=\widetilde{F}_{\mu\nu}F^{\mu\nu}$.

The energy-momentum tensor associated with the Bumblebee field is explicitly provided by
\begin{equation}
\begin{aligned}
&\mathcal{T}^{\rm B}_{\mu\nu} = \xi\biggr\{
   \tfrac{1}{2}B^{\alpha}B^{\beta}R_{\alpha\beta}\,g_{\mu\nu}
   - B_{\mu}B^{\alpha}R_{\alpha\nu}
   - B_{\nu}B^{\alpha}R_{\alpha\mu} \\
   &+ \tfrac{1}{2}\nabla_{\alpha}\nabla_{\mu}\bigl(B^{\alpha}B_{\nu}\bigr)
   + \tfrac{1}{2}\nabla_{\alpha}\nabla_{\nu}\bigl(B^{\alpha}B_{\mu}\bigr)
    - \tfrac{1}{2}\nabla^{2}\bigl(B_{\mu}B_{\nu}\bigr)
   \\
& - \tfrac{1}{2}\,g_{\mu\nu}\,\nabla_{\alpha}\nabla_{\beta}\bigl(B^{\alpha}B^{\beta}\bigr)
\biggl\}+ 2V'\,B_{\mu}B_{\nu}
+ B_{\mu}{}^{\alpha}B_{\nu\alpha}
\\
&- \Bigl(V + \tfrac{1}{4}B_{\alpha\beta}B^{\alpha\beta}\Bigr)\,g_{\mu\nu}
- \tfrac{1}{2}\,g_{\mu\nu}\Bigl[
   \xi\bigl(
     \tfrac{1}{2}B^{\alpha}B^{\beta}R_{\alpha\beta}\,g^{\rho\sigma}
     \\
     &- B_{\rho}B^{\alpha}R_{\alpha}{}^{\rho}
 - B_{\sigma}B^{\alpha}R_{\alpha}{}^{\sigma}\\
        &+ \tfrac{1}{2}\nabla_{\alpha}\nabla^{\rho}\bigl(B^{\alpha}B_{\rho}\bigr)
     + \tfrac{1}{2}\nabla_{\alpha}\nabla^{\sigma}\bigl(B^{\alpha}B_{\sigma}\bigr)
     - \tfrac{1}{2}\nabla^{2}\bigl(B_{\lambda}B^{\lambda}\bigr)
     \\&- \tfrac{1}{2}\,g_{\rho\sigma}g^{\rho\sigma}\,
         \nabla_{\alpha}\nabla_{\beta}\bigl(B^{\alpha}B^{\beta}\bigr)
     + 2V'\,B_{\lambda}B^{\lambda}
   \bigr)
\Bigr].
\end{aligned}
\end{equation}

By varying the action \eqref{S} with respect to the bumblebee vector field $B_\mu$ and the electromagnetic potential $A_\mu$, we obtain the corresponding equations of motion, which encapsulate the coupled dynamics of gravity, nonlinear electrodynamics, and spontaneous Lorentz-symmetry breaking. This framework enables a systematic analysis of
\begin{equation}
\begin{aligned}
	&\nabla_{\mu} B^{\mu\nu} - 2\bigg( V' B^\nu - \frac{\xi}{2} B_\mu R^{\mu\nu}- \frac{1}{2} \eta B^\nu F^{\alpha\beta} F_{\alpha\beta} \bigg) \\
    & - \frac{1}{2} \cosh\gamma B^\nu F_{\mu\nu} F^{\mu\nu}  \\
&+ \frac{1}{2}  \sinh\gamma B^\nu \left[ \left( \frac{1}{4} F_{\mu\nu} F^{\mu\nu} \right)^2 + \left( \frac{1}{4} F_{\mu\nu} \widetilde{F}^{\mu\nu} \right)^2 \right] = 0,\label{BB}
\end{aligned}
\end{equation}

\begin{equation}
\begin{aligned}
&\nabla^\nu \Biggr\{  \Bigg( \left( -\cosh\gamma + \frac{1}{4} F_{\mu\nu} F^{\mu\nu} \sinh\gamma \right) F_{\mu\nu} \\
&+ \frac{1}{4} F_{\mu\nu} \tilde{F}^{\mu\nu} \sinh\gamma \, \tilde{F}_{\mu\nu} \Bigg) (1 + \eta B^{\alpha} B_{\alpha}) \Biggl\}
=0.
	\label{FF}
\end{aligned}
\end{equation}

To investigate the black hole solution in more detail, we restrict our attention 
to the electrically charged case. We impose the ansatz $A_{\mu} = \Phi(r)\,\delta_{\mu}^{0}$, where $\Phi(r)$ denotes the electrostatic potential. This choice reduces the 
vector potential to a single radial function, thereby simplifying the analysis 
and allowing us to focus on the essential features of the electromagnetic field.

\section{Black Hole Solutions} \label{sec2}
\subsection{Static, Spherically Symmetric Black Hole}
\noindent
To investigate electrically charged black hole solutions in the context of bumblebee gravity coupled to traceless-conformal electrodynamics field ``ModMax'', we consider a static, spherically symmetric spacetime described by the line element
\begin{equation}
\mathrm{d}s^2= -\mathcal{G}_1(r)\,\mathrm{d}t^2 + \frac{\mathrm{d}r^2}{\mathcal{G}_2(r)} + r^2 \left(\mathrm{d}\theta^2 + \sin^2\theta\,\mathrm{d}\varphi^2\right),
	\label{qdc}
\end{equation}
where $\mathcal{G}_1(r)$ and $\mathcal{G}_2(r)$ are the lapse and radial metric functions, respectively, to be determined from the coupled field equations.  

Following the procedure of Ref.~\cite{1LIV2}, we assume that the bumblebee vector field $B_\mu$ acquires a nonzero vacuum expectation value $b_\mu$, thereby spontaneously breaking local Lorentz symmetry. Motivated by previous studies~\cite{Bertolami:2005bh}, we take this vacuum configuration to be purely radial:
\begin{equation}
	b_\mu = \bigl(0,\, b_r(r),\, 0,\, 0 \bigr),
\end{equation}
and enforce the constraint that the vector field maintains a fixed norm,
\begin{equation}
	b_\mu b^\mu = b^2 = \text{const},
\end{equation}
which immediately implies
\begin{equation}
	b_r(r) = b \,\sqrt{\mathcal{G}_2(r)}.
\end{equation}

This ansatz ensures compatibility with the spacetime symmetries while encoding the spontaneous Lorentz violation in the radial direction, thereby reducing the bumblebee sector to a single function $b_r(r)$ that directly couples to the metric and the nonlinear electromagnetic field. The resulting setup provides a tractable framework for exploring the combined effects of Lorentz symmetry breaking and ModMax nonlinearity on the structure and horizons of electrically charged black holes.

To rigorously investigate the interplay between ModMax electrodynamics and spontaneous Lorentz symmetry breaking, we introduce the Lorentz-violating parameter
$\ell = \xi\, b^2$ \cite{Liu:2024axg}.
Considering the static, spherically symmetric metric \eqref{qdc}, the modified field equations can expressed as follows
\begin{equation}
\begin{aligned}
&\frac{2+\ell}{8\,\mathcal{G}_2(r)}\Bigl(\frac{\mathcal{G}_1'(r)^2}{\mathcal{G}_1(r)} + 2\,\mathcal{G}_1''(r)\Bigr)
- \frac{2+\ell}{8\,\mathcal{G}_2(r)^2}\,\mathcal{G}_1'(r)\,\mathcal{G}_2'(r)
\\&- \frac{1+\ell}{r\,\mathcal{G}_2(r)}\,\mathcal{G}_1'(r)- \frac{\ell}{2\,r\,\mathcal{G}_2(r)^2}\,\mathcal{G}_1(r)\,\mathcal{G}_2'(r)\\
&+\Bigl[\Lambda +\bigl(V'(X)\,b^2 + V(X)\bigr)\Bigr]\,\mathcal{G}_1(r)
\\&-\frac{1+2\,b^2\eta}{2\mathcal{G}_2(r)}\bigl(\cosh\gamma + \sinh\gamma\bigr)\,\Phi'(r)^2
=\;0,\qquad \label{G1}
\end{aligned}
\end{equation}
\begin{equation}
\begin{aligned}
&\frac{\mathcal{G}_1'(r)\,\mathcal{G}_2'(r)}{4\,\mathcal{G}_1(r)\,\mathcal{G}_2(r)}
-\frac{\mathcal{G}_1''(r)}{2\,\mathcal{G}_1(r)}
+\frac{\mathcal{G}_1'(r)^2}{4\,\mathcal{G}_1(r)^2}
+\frac{\mathcal{G}_2'(r)}{r\,\mathcal{G}_2(r)}
\\
&-\frac{(1+2\,b^2\eta)\,(\cosh\gamma + \sinh\gamma)\,\Phi'(r)^2}{(2+3\ell)\,\mathcal{G}_1(r)}\\
&-\frac{2\,\Lambda\,\mathcal{G}_2(r)}{2+3\ell}
-2\,\frac{\bigl(V'\,b^2 + V\bigr)\,\mathcal{G}_2(r)}{2+3\ell}
\;=\;0,\, \label{G2}
\end{aligned}
\end{equation}
\begin{equation}
\begin{aligned}
& 1 - \frac{1+\ell}{\mathcal{G}_2} - \Lambda\,r^2 -\,(V'b^2 + V)\,r^2  + \frac{r}{2\,\mathcal{G}_2^2}\,\mathcal{G}_2' 
   \\&- \frac{\ell\,r^2}{8\,\mathcal{G}_1\,\mathcal{G}_2^2}\,\mathcal{G}_1'\,\mathcal{G}_2' 
   - \frac{\ell\,r^2}{8\,\mathcal{G}_1^2\,\mathcal{G}_2}\,\mathcal{G}_1'^2 - \frac{(1+\ell)\,r}{2\,\mathcal{G}_1\,\mathcal{G}_2}\,\mathcal{G}_1'
    \\
&+ \frac{\ell\,r^2}{4\mathcal{G}_1\mathcal{G}_2}\,\mathcal{G}_1'' - \frac{(1+2b^2\eta)(\cosh\gamma + \sinh\gamma)}{2}r^2\Phi'^2 =0, \label{G3}
\end{aligned}
\end{equation}
\begin{equation}
\begin{aligned}
	&\bigg( \frac{r^2}{\sqrt{\mathcal{G}_1(r) \mathcal{G}_2(r)}} \left( \cosh\gamma + \sinh\gamma \right) F_{tr} \bigg)_{,r} =0.\label{fff}
\end{aligned}
\end{equation}

Our focus is on discussing plausible solutions for the field equations, with the aim of addressing a physical black hole solution within the framework of the imposed assumptions.

In the absence of a cosmological constant, and following the framework of Casana et al.~\cite{1LIV2}, we impose the vacuum constraints 
$V=0$ and $V'=0$, which guarantee that the self-interaction potential does not contribute explicitly to the field equations. A particularly natural choice that realizes these conditions is the smooth quadratic form
\begin{equation}
V(X) = \frac{\lambda}{2}X^{2},\label{iia}
\end{equation}
with $\lambda$ denoting a coupling constant. Such a potential is well known from the paradigm of spontaneous symmetry breaking, most prominently in the Higgs mechanism~\cite{Higgs:1964ia,Englert:1964et}, and acquires a deeper significance within Lorentz-violating extensions of gravity~\cite{Kostelecky:1989jw,Kostelecky:2004hs}. In the bumblebee framework, the vector field develops a nonvanishing vacuum expectation value that spontaneously breaks local Lorentz symmetry, thereby endowing the background spacetime with preferred directions. The quadratic potential plays a central role in stabilizing this vacuum configuration while leaving the classical dynamics of the field equations unaltered under the imposed constraints.

However, higher-order deformations, such as 
\begin{equation}
V(X) = \frac{\lambda}{2}X^{n}, \quad n \geq 3,
\end{equation}
also satisfy the same vacuum requirements, the quadratic model remains the simplest and most analytically tractable realization. Moreover, it captures the essential physics of vacuum alignment and mass generation for excitations about the Lorentz-violating ground state. Thus, within bumblebee gravity, the quadratic potential not only provides technical simplification but also embodies the minimal and most effective mechanism for encoding controlled Lorentz-symmetry breaking while preserving the overall consistency of the theory.

Leveraging Eq.~\eqref{fff}, the radial electric field component can be expressed as
\begin{equation}
F_{tr} = \sqrt{\mathcal{G}_1(r)\,\mathcal{G}_2(r)}\,\phi'(r).
\end{equation}

Substituting this relation for $F_{tr}$, together with the quadratic potential $V(X)$ from Eq.~\eqref{iia}, into the previously established constraints, and adopting the coupling identification 
\begin{equation}
\eta = \frac{\xi}{2+\ell},
\end{equation} 
we obtain the following closed set of modified field equations governing the coupled bumblebee–ModMax system:

\begin{equation}
\begin{aligned}
&\frac{2+\ell}{8\,\mathcal{G}_2(r)}\Bigl(\frac{\mathcal{G}_1'(r)^2}{\mathcal{G}_1(r)} + 2\,\mathcal{G}_1''(r)\Bigr)
- \frac{2+\ell}{8\,\mathcal{G}_2(r)^2}\,\mathcal{G}_1'(r)\,\mathcal{G}_2'(r)\\
&- \frac{1+\ell}{r\,\mathcal{G}_2(r)}\,\mathcal{G}_1'(r) - \frac{\ell}{2\,r\,\mathcal{G}_2(r)^2}\,\mathcal{G}_1(r)\,\mathcal{G}_2'(r)  \\
&- \frac{1 + 2 \frac{\ell}{2 + \ell}}{2\mathcal{G}_2(r)} \bigl(\cosh\gamma + \sinh\gamma\bigr) \Phi'(r)^2 = 0,\label{S1} \qquad
\end{aligned}
\end{equation}
\begin{equation}
\begin{aligned}
&\frac{\mathcal{G}_1'(r)\,\mathcal{G}_2'(r)}{4\,\mathcal{G}_1(r)\,\mathcal{G}_2(r)} - \frac{\mathcal{G}_1''(r)}{2\,\mathcal{G}_1(r)} + \frac{\mathcal{G}_1'(r)^2}{4\,\mathcal{G}_1(r)^2} + \frac{\mathcal{G}_2'(r)}{r\,\mathcal{G}_2(r)}\\
&- \frac{(1 + 2\,\frac{\ell}{2 + \ell})(\cosh\gamma + \sinh\gamma) \Phi'(r)^2}{(2 + 3\ell)\,\mathcal{G}_1(r)} = 0,\label{S2} \quad 
\end{aligned}
\end{equation}
\begin{equation}
\begin{aligned}
&1 - \frac{1+\ell}{\mathcal{G}_2(r)} + \frac{r}{2\,\mathcal{G}_2^2}\,\mathcal{G}_2' 
+ \frac{\ell\,r^2}{4\,\mathcal{G}_1\,\mathcal{G}_2}\,\mathcal{G}_1'' - \frac{\ell\,r^2}{8\,\mathcal{G}_1\,G^2}\,\mathcal{G}_1'\,\mathcal{G}_2' 
\\&- \frac{\ell\,r^2}{8\,\mathcal{G}_1^2\,\mathcal{G}_2}\,\mathcal{G}_1'^2 - \frac{(1+\ell)\,r}{2\,\mathcal{G}_1\,\mathcal{G}_2}\,\mathcal{G}_1' \\
&- \frac{(1+2\,\frac{\ell}{2+\ell})\,(\cosh\gamma + \sinh\gamma)}{2}\,r^2\,\Phi'^2 = 0,\quad\label{S3} 
\end{aligned}
\end{equation}
\begin{equation}
\begin{aligned}
&\Phi'(r) \bigg(  \frac{r^2\left( \mathcal{G}_1'(r) \mathcal{G}_2(r) + \mathcal{G}_1(r) \mathcal{G}_2'(r) \right)}{2 \sqrt{\mathcal{G}_1(r) \mathcal{G}_2(r)}^3}-\frac{2r}{\sqrt{\mathcal{G}_1(r) \mathcal{G}_2(r)}}   \bigg) \\&- \frac{r^2}{\sqrt{\mathcal{G}_1(r) \mathcal{G}_2(r)}} \Phi''(r) =0.\label{S4}
\end{aligned}
\end{equation}

By simplifying the field equations \eqref{S1} and \eqref{S2}, one finds the relation
\begin{equation}
\frac{d}{dr} \Bigl[ \mathcal{G}_2(r)\, \mathcal{G}_1(r) \Bigr] = 0,
\end{equation}
which, upon integration, yields
\begin{equation}
\mathcal{G}_2(r) = \frac{C_1}{\mathcal{G}_1(r)}.
\end{equation}
Following the prescription of Casana et al.~\cite{1LIV2}, we fix the integration constant as $C_1 = 1 + \ell$, so that the limit $\ell \to 0$ smoothly recovers the standard Schwarzschild normalization, while a nonzero $\ell$ parametrizes the leading-order Lorentz-violating deformation of the spacetime geometry.

Within our static, spherically symmetric ansatz, the only nonvanishing component of the electromagnetic field strength is the radial electric field, $E(r) = F_{tr}$. 
Consequently, the full bumblebee–ModMax black hole solution can be compactly expressed as 
\begin{equation}
\mathcal{G}_2(r) = \frac{1+\ell}{\mathcal{G}_1(r)}, \qquad \phi(r) = \frac{e^{\gamma} Q_0}{r},
\end{equation}
where the Lorentz-violating parameter $\ell$ and the nonlinear ModMax deformation $\gamma$ appear explicitly, highlighting their respective effects on spacetime geometry and electromagnetic field.  

The exact metric and electrostatic potential are therefore given by
\begin{align}
\mathcal{G}_1(r) =\frac{1+\ell}{\mathcal{G}_2(r)}= 1 - \frac{2M}{r} + \frac{2 (1+\ell) Q_0^2}{(2+\ell) r^2}.\label{Q11}
\end{align}

This solution clearly demonstrates how the Lorentz-violating shift $\ell$ modifies the gravitational potential, effectively rescaling the contribution of the electric charge, while the ModMax parameter $\gamma$ induces a nonlinear deformation of the Coulomb field. The resulting spacetime generalizes the standard Reissner–Nordström geometry to include controlled Lorentz-violating and nonlinear electrodynamic effects.

Next, the conserved current acquires an additional contribution from the bumblebee coupling, taking the form
\begin{equation}
J^{\mu} = - \left(\cosh\gamma + \sinh\gamma\right) \nabla_{\nu} \Bigl( F^{\mu\nu} + \eta B^{\alpha} B_{\alpha} F^{\mu\nu} \Bigr),
\end{equation}
where the field strength is purely radial, $F_{\mu\nu} = -\phi_{,r}\, \delta^0_{[\mu} \delta^1_{\nu]}$~\cite{Flores-Alfonso:2020euz}.  

The total electric charge $Q$ can then be expressed as a flux integral over a spacelike hypersurface $S^2_\infty$ at spatial infinity, via Stokes's theorem~\cite{Carroll2019}:
\begin{equation}
\begin{aligned}
Q &= - \frac{1}{4\pi} \int_{S^2_\infty} d^3x \, \sqrt{\gamma^{(3)}} \, n_\mu J^\mu  \\
&= \frac{1}{4\pi} \int_{\partial S^2_\infty} d\theta\, d\phi \, \sqrt{\gamma^{(2)}} \, n_\mu \sigma_\nu 
\,e^\gamma \left( F^{\mu\nu} + \frac{\xi B^\alpha B_\alpha F^{\mu\nu}}{\ell+2}  \right)  \\
&= \left(1 + b^2 \frac{\xi}{\ell+2}\right) \left( \cosh\gamma + \sinh\gamma \right) Q_0  \\
&= \frac{2(1+\ell)}{2+\ell} \left( \cosh\gamma + \sinh\gamma \right) Q_0.
\end{aligned}
\end{equation}

Here, $n_\mu = (1,0,0,0)$ is the unit normal to $S^2_\infty$ with the induced metric $\gamma^{(3)}_{ij}$, while $\sigma_\mu = (0,1,0,0)$ is the outward normal on the boundary two-sphere $\partial S^2_\infty$, whose induced metric reads $\gamma^{(2)}_{ij} = r^2 (d\theta^2 + \sin^2\theta\, d\phi^2)$.  

The electrically charged ModMax black hole in bumblebee gravity is therefore fully characterized by the metric functions
\begin{align}\label{eq39}
\mathcal{G}_1(r) &= 1 - \frac{2M}{r} + \frac{e^{-\gamma} (2+\ell) Q^2}{2 (1+\ell) r^2}, \\
\mathcal{G}_2(r) &= \frac{1+\ell}{\mathcal{G}_1(r)},
\end{align}
which encapsulate the combined effects of the Lorentz-violating shift $\ell$ and the nonlinear ModMax deformation $\gamma$ on the spacetime geometry.

The dimensionless parameter $\gamma$ plays a crucial role in ModMax electrodynamics by determining the relative weight of the two conformal invariants. It effectively modifies the Coulomb term, either suppressing or enhancing it through factors of $e^{\pm\gamma}$. Consequently, the electromagnetic sector induces a deformation of the spacetime geometry relative to the standard Reissner–Nordström form. In the simultaneous limit $(\gamma,\eta\to0)$, the solution continuously reduces to the familiar Reissner–Nordström metric of Einstein–Maxwell theory. A nonvanishing $\eta$. representing the strength of the bumblebee coupling, introduces controlled Lorentz-violating corrections to the effective electric energy density, thereby shifting the horizon structure and altering the causal properties of the geometry in characteristic ways~\cite{1LIV1}. Notably, in contrast to numerous charged black hole solutions in modified gravity that simultaneously deform both the $M/r$ and $Q^2/r^2$ terms or introduce nontrivial radial dependencies as occurs, for example, in Einstein–Gauss–Bonnet gravity~\cite{Fernandes:2020rpa}, Eddington-inspired Born–Infeld gravity~\cite{Wei:2014dka}, or Einstein–Maxwell–\ae{}ther gravity~\cite{Ding:2015kba}, our framework induces a more controlled modification. Specifically, the nonlinear ModMax electrodynamics and bumblebee-induced Lorentz violation act primarily to rescale the Coulomb term according to
\begin{equation}
\frac{e^{-\gamma}\,(2+\ell)}{2\,(1+\ell)},
\end{equation}
while leaving the mass term in the canonical Schwarzschild form, $-2M/r$, unaltered.  

In the asymptotic regime, the metric functions approach
\begin{equation}
\mathcal{G}_1(r) \to 1, \qquad \mathcal{G}_2(r) \to 1+\ell,
\end{equation}
indicating that the spacetime does not approach exact Minkowski space but rather a Lorentz-shifted vacuum with $\mathcal{G}_{2\infty} = 1+\ell$. This asymptotic structure reflects the leading-order imprint of spontaneous Lorentz-symmetry breaking, while the nonlinear ModMax deformation $\gamma$ selectively modifies the Coulombic contribution, thereby preserving the mass term and simplifying the horizon structure relative to more intricate modified-gravity scenarios.

\subsection{Slowly Rotating Black Hole}
In 1963 Kerr derived the exact solution describing a rotating (vacuum) black hole~\cite{Kerr:1963ud}, and shortly thereafter Newman and collaborators extended this construction to include electric charge, yielding the Kerr--Newman family~\cite{Newman:1965my}. Rotating solutions have also been obtained in the context of bumblebee gravity (see, e.g., \cite{Ding:2019mal,Ding:2020kfr,Jha:2020pvk}).

In this work we study slowly rotating black hole configurations in bumblebee gravity with a traceless conformal electrodynamics in the absence of a cosmological constant. 
We adopt the following general ansatz for the metric:
\begin{equation}
\begin{aligned}
\label{metric:ansatz}
ds^2 =& -A(r,\theta)\,dt^2 + S(r,\theta)\,dr^2 + 2\,F(r)\,H(\theta)\,a\,dt\,d\phi\\
&+ \rho(r,\theta)^2\,d\theta^2 + h(r,\theta)^2\sin^2\theta\,d\phi^2,
\end{aligned}
\end{equation}
and take the bumblebee and electromagnetic fields to have the form
\begin{align}
\label{bumblebee:field}
&b_{\mu} = \bigl(0,\; b_r(r,\theta),\; 0,\; 0\bigr),\\
\label{em:field}
&F_{\mu\nu} = \partial_{\mu}A_{\nu}-\partial_{\nu}A_{\mu},\\
\label{A:ansatz}
&A_{\mu} = \bigl(A_0(r,\theta),\;0,\;0,\;A_{\phi}(r,\theta)\bigr).
\end{align}

Consistency with known limits constrains the solution: in the vanishing Lorentz-violating limit the geometry must reduce to the Kerr-Newman spacetime, while for zero rotation parameter $a$ one recovers the spherically symmetric black hole obtained in the previous section. 
Accordingly, we treat the product $a\ell$ as a small parameter and construct the solution perturbatively in the slow-rotation regime. 
Building on the slowly rotating constructions in bumblebee gravity of Ding \emph{et al.} \cite{Ding:2019mal,Ding:2020kfr}, we derive a slowly rotating black hole solution coupled to ModMax field within the bumblebee framework valid to leading order in the rotation parameter.

We generalize to a rotating, axially symmetric spacetime by adopting a Kerr‐like ansatz in Boyer–Lindquist coordinates $(t,r,\theta,\phi)$ \cite{Kerr:1963ud,Boyer:1966qh,Carter:1968ks}:
\begin{equation}
\begin{aligned}
\label{eq:rotatingMetric}
\!\!\!ds^2 &=\!-\!\frac{\Delta_r}{\rho^2}\Bigl(dt - a\sqrt{1+\ell}\,\sin^2\theta\,d\phi\Bigr)^2
   + (1+\ell)\,\frac{\rho^2}{\Delta_r}\,dr^2 \\
&\!+\! \rho^2\,d\theta^2   + \frac{\sin^2\theta}{\rho^2}\Bigl(a\sqrt{1+\ell}\,dt 
   - \bigl(r^2 + a^2(1+\ell)\bigr)d\phi\Bigr)^2 \,,
\end{aligned}
\end{equation}
with
\begin{align}
\label{eq:DeltaR}
\Delta_r &= r^2 + a^2(1+\ell) - 2Mr 
   + \frac{e^{-\gamma}\,(\ell+2)\,Q^2}{2(1+\ell)}\\
   \rho^2 &= r^2 + a^2(1+\ell)\cos^2\theta\,. 
\end{align}
The nonzero components of the Maxwell-ModMax potential and the bumblebee radial field are
\begin{align}
\label{eq:VectorPotentials}
&A_t(r,\theta)  = -\,\frac{e^{-\gamma}\,Q\,r}{\rho^2}\,, 
\\
&A_\phi(r,\theta) = \frac{e^{-\gamma}\,Q\,a\,\sqrt{1+\ell}\,r\sin^2\theta}{\rho^2}\,, 
\\
\label{eq:BumblebeeField}
&b_r(r) = b\,\rho\,\sqrt{\frac{1+\ell}{\Delta_r}}\,,
\end{align}
where $a$ is the rotation parameter and $b$ sets the overall scale of the bumblebee vacuum expectation value.

Horizons are located at the roots of $\Delta_r=0$:
\begin{equation}
\label{eq:Horizons}
r_{\pm} = M \pm \sqrt{\,M^2 \;-\; a^2(1+\ell) \;-\;\frac{e^{-\gamma}(\ell+2)\,Q^2}{2(1+\ell)}\,}\,.
\end{equation}
In the appropriate limits, our solution reduces to familiar geometries: for 
$\ell \to 0$ and $\gamma \to 0$, it smoothly reproduces the standard 
Kerr-Newman spacetime with horizons 
\begin{equation}
r_{\pm} = M \pm \sqrt{M^{2} - a^{2} - Q^{2}}\, .
\end{equation}
A nonzero Lorentz-violation parameter $\ell$ shifts the asymptotic frame, such that 
\begin{equation}
\frac{g_{\phi\phi}}{\sin^{2}\theta} \;\to\; r^{2} + a^{2}(1+\ell)
\quad \text{as} \quad r \to \infty \, .
\end{equation}
Meanwhile, the ModMax deformation parameter $\gamma$ effectively rescales the 
electric charge according to $Q \to e^{-\gamma/2} Q$, thereby modifying the 
Coulombic contribution in $\Delta_{r}$. The bumblebee radial profile,
\begin{equation}
b_{r}(r) = b\,\rho\,\sqrt{\frac{1+\ell}{\Delta_{r}}},
\end{equation}
remains real for all $r \geq r_{+}$, guaranteeing a regular Lorentz-violating background outside the outer horizon. This compact form provides a convenient  basis for subsequent analyses of massive scalar QNMs properties in the slowly  rotating bumblebee--ModMax spacetime.

To ensure that this slow rotation ansatz effectively solves the complete Einstein-Bumblebee ModMax system at $\mathcal O(a)$, we incorporate the tensor
\begin{equation}
\begin{aligned}
\!\!\!\Delta_{\mu\nu}\!-\! R_{\mu\nu}\! =&  - \Biggr\{V'\bigl(2B_{\mu}B_{\nu} + b^{2}g_{\mu\nu}\bigr)
+ B_{\mu}{}^{\alpha}B_{\nu\alpha}\\
&+ V\,g_{\mu\nu}\tfrac14 B_{\alpha\beta}B^{\alpha\beta}\Biggl\}
 - \bigl(T^{\rm M}{}_{\mu\nu} - \tfrac12g_{\mu\nu}T^{\rm M}\bigr)\\
&- \xi\Biggr\{\tfrac12B^{\alpha}B^{\beta}R_{\alpha\beta}g_{\mu\nu}
  \!-\! \tfrac12\nabla^{2}(B_{\mu}B_{\nu}) \!-\! B_{\nu}B^{\alpha}R_{\alpha\mu}\\
&-\! B_{\mu}B^{\alpha}R_{\alpha\nu}\!+\! \tfrac12\nabla_{\alpha}\nabla_{\nu}(B^{\alpha}B_{\mu}) \!+\! \tfrac12\nabla_{\alpha}\nabla_{\mu}(B^{\alpha}B_{\nu})
\Biggl\}.
\end{aligned}
\end{equation}
Obviously, the full set of field equations is satisfied (finite $Q)$, one requires $\Delta_{\mu\nu}=0$.  Substituting our metric and gauge-field expansions then yields
\begin{equation}
\Delta_{\mu\nu} =
\begin{cases}
\displaystyle \mathcal O(a^2\,\ell)\,, 
& (\mu\nu)=tt,\,rr,\,\theta\theta,\,\phi\phi,\,r\theta,\,t\phi,\\
0\,, & \text{otherwise.}
\end{cases}\label{sloww}
\end{equation}
Thus at first order in $a$, every component of the rotating solution satisfies the full system up to suppressed $\mathcal O(a^2\ell)$ corrections, imposing no further constraints on $\{M,Q,\ell,\gamma\}$ beyond the static extremality bound of Sect. \ref{sec2}.

\section{MASSIVE SCALAR PERTURBATIONS OF SLOWLY ROTATING  BLACK HOLES} \label{sec3}
\newcommand{\pp}{\partial}

\subsection{Massive scalar  Perturbation equation}

Within the first-order slow-rotation approximation, the rotating spacetime obtained above can be conveniently employed to investigate how parameters and spin affect the dynamical evolution of scalar fields, which in turn serves to evaluate its physical consistency.
By defining the dimensionless spin parameter $\tilde{a}=a/M$ and retaining terms to first order in a Taylor expansion, we obtain the following line element for the metric:
\begin{equation}
\begin{aligned}
 ds^2_{(1)}=&-\left(1-\frac{2M}{r}+\frac{e^{-\gamma}Q^2(2+\ell)}{2r^2(1+\ell)} \right)dt^2 \\ 
 &+(1+\ell)\left(1-\frac{2M}{r}+\frac{e^{-\gamma}Q^2(2+\ell)}{2r^2(1+\ell)} \right)^{-1}dr^2 \\ 
 &+r^2d\theta^2+r^2\sin^2\theta d\phi^2-2 \tilde{a} M\sqrt{1+\ell}\sin^2\theta \\
 &\times\left( \frac{2M}{r}-\frac{e^{-\gamma}Q^2(2+\ell)}{2r^2(1+\ell)} \right)dt d\phi+\mathcal{O}(\tilde{a}^2).
\end{aligned}
\end{equation}
At this order, the event horizon $ r_h $ and the Cauchy horizon $ r_m $ can be expressed as follows:
\begin{equation}
\begin{aligned}
r_h=M+M\sqrt{1-\frac{e^{-\gamma}Q^2(2+\ell)}{2M^2(1+\ell)}},\\
\quad\quad r_m=M-M\sqrt{1-\frac{e^{-\gamma}Q^2(2+\ell)}{2M^2(1+\ell)}}.
\end{aligned}
\end{equation}
For this, we can rewrite the line element using $r_h$ and $r_m$ as:
\begin{equation}
\begin{aligned}
\!\!\! ds^2_{(1)}=&-F(r)dt^2+\frac{(1+\ell)}{F(r)}dr^2+r^2d\theta+r^2\sin^2\theta d\phi^2\\
&-2 \tilde{a} M\sqrt{1+\ell}\sin^2\theta \left( \frac{2M}{r}-\frac{e^{-\gamma}Q^2(2+\ell)}{2r^2(1+\ell)} \right)dt d\phi,
\end{aligned}
\end{equation}
where
\begin{equation}
F(r)=\left(1-\frac{r_h}{r}\right)\left(1-\frac{r_m}{r}\right).
\end{equation}
In the absence of charge ($Q=0$), our results are consistent with those for slow-rotation Lorentz-violating black holes reported by Ding et al. \cite{2.281}. 
However, it is worth noting that when the non-linearity parameter $\gamma=0$, the solution does not reduce to the charged bumblebee black hole; this is a characteristic feature of Conformal Nonlinear Electrodynamics.

Considering that the scalar field does not couple directly to the bumblebee field, the dynamics of a massive scalar field are governed by the Klein–Gordon equation
\begin{equation}
\frac{1}{\sqrt{-g}}\pp_\mu \left( \sqrt{-g}g^{\mu\nu}\pp_\nu \varphi \right)=\mu^2\varphi,
\end{equation}
where $m_s = \mu \hbar$ denotes the scalar field mass.

To separate variables, we expand the scalar field in terms of spherical harmonics,
\begin{equation}
\varphi(t,r,\theta,\phi) = \sum_{l m} \frac{\Psi(t,r)}{r} \, Y^{lm}(\theta,\phi),
\end{equation}
and subsequently perform an expansion with respect to the dimensionless spin parameter $\tilde{a}$. Making use of the eigenvalue equation for the angular functions,
\begin{equation}
\left[\frac{1}{\sin\theta}\frac{\pp}{\pp\theta}\left(\sin\theta \frac{\pp}{\pp \theta} \right)
+\frac{1}{\sin^2\theta}\frac{\pp^2}{\pp \phi^2}\right]Y^{lm}=l(l+1)Y^{lm},
\end{equation}
the Klein–Gordon equation can then be reduced to an effective radial equation as
\begin{equation}
\begin{aligned}
F(r)\frac{\pp^2}{\pp r^2}\Psi(t,r)-\frac{(1+\ell)}{F(r)}\frac{\pp^2}{\pp^2_t}\Psi(t,r)+F'(r)\Psi(t,r)  \\
-\frac{2imM\tilde{a}(1+\ell)^{3/2}}{r^2F(r)}\left(\frac{2M}{r}-\frac{e^{-\gamma}Q^2(2+\ell)}{2r^2(1+\ell)}\right)\frac{\pp}{\pp t}\Psi(t,r)\\
-\left[\frac{l^2+l+r^2\mu^2}{r^2(1+\ell)^{-1}}+\frac{2Mr-e^{-\gamma}Q^2\left(\tfrac{2+\ell}{1+\ell}\right)}{r^4(1+\ell)^2}\right] \Psi(t,r)=0.
\end{aligned}
\end{equation}

To compute the eigenfrequencies, we work in the frequency domain. 
Assuming a harmonic time dependence $e^{-i\omega t}$, i.e., $\Psi(t,r)=e^{-i\omega t}\psi(r)$, the foregoing equation can be recast into a Schrödinger-like form:
\begin{equation}\label{eq111}
\left[\frac{d^2}{dr_*^{2}}+(1+\ell)\omega^{2}- \mathcal{V}_l \right]\psi_{l}=0,
\end{equation}
where the tortoise coordinate is defined as
\begin{equation}
r_*=\int F(r)^{-1}dr=r+\frac{r_h^2\ln\left(r-r_h\right)}{r_h-r_m}-\frac{r_m^2\ln\left(r-r_m\right)}{r_h-r_m},
\end{equation}
and $\mathcal{V}_l$ represents the effective potential of the field in the slow-rotation expansion:
\begin{equation}
\begin{aligned}\label{eq69}
\mathcal{V}_l=&(1+\ell)F\left(\frac{l^2+l+r^2\mu^2}{r^2}+\frac{2Mr-e^{-\gamma}Q^2\left(\tfrac{2+\ell}{1+\ell} \right)}{r^4(1+\ell)}  \right)\\
&+\frac{mM\tilde{a}\omega\sqrt{1+\ell}}{r^2e^{\gamma}}\left( \frac{4e^{\gamma}M(1+\ell)}{r}-
\frac{Q^2(2+\ell)}{r^2} \right).
\end{aligned}
\end{equation}
Having recast the perturbation equation into this form, one may impose the appropriate boundary conditions at the horizon and at spatial infinity. The resulting eigenvalue problem can then be solved using standard numerical techniques to extract the QNM spectrum.

\subsection{Boundary conditions}

We aim to provide a comprehensive investigation of the QNMs of a massive scalar field in rotating bumblebee black hole spacetimes endowed with conformal nonlinear electrodynamics. 
Prior to this work, Ref. \cite{Hu:2025isj} examined massive-field perturbations within this gravity model, and Ref. \cite{Deng:2025uvp} subsequently extended the analysis to spinning configurations, discussing two distinct rotating bumblebee black holes. 
However, the implications specific to conformal nonlinear electrodynamics in the rotating case remain largely unexplored.
To solve Eq. \meq{eq111}, we first extract its asymptotic solutions at the two boundaries.
Accordingly, we expand the radial equation near the event horizon $r_h$ and at spatial infinity, obtaining, respectively,
\begin{equation}\label{eq211}
\psi_l''(r_*)+\left[(1+\ell)\omega^2-\mathcal{V}_H\right] \psi_l(r_*)=0,
\end{equation}
with 
\begin{equation}
\mathcal{V}_H=\frac{(1+\ell)^{3/2}mM\tilde{a}\omega}{r_h^3}\left[ 4M-\frac{e^{-\gamma}Q^2(2+\ell)}{r_h(1+\ell)} \right],
\end{equation}
and
\begin{equation}\label{eq212}
\psi_l''(r_*)+(1+\ell)\left(\omega^2-\mu^2\right) \psi_l(r_*)=0.
\end{equation}

Both Eq.~\meq{eq211} and Eq.~\meq{eq212} admit two independent solutions. To select the physically relevant modes, we impose appropriate boundary conditions: only ingoing waves are allowed at the event horizon, while only outgoing waves are permitted at spatial infinity. Under these conditions, the asymptotic behavior of the wave function is given by
\begin{equation}
\psi_l \sim
\begin{cases}
e^{-i\sqrt{1+\ell}K_Hr_*}, & \text{for } r_* \rightarrow -\infty, \\
e^{i\sqrt{(1+\ell)q^2}\,r_*}, & \text{for } r_* \rightarrow +\infty .
\end{cases}
\end{equation}
Here, the horizon wave number is
\begin{equation}
\begin{aligned}
K_H=\sqrt{\mathcal{V}_H}
=\sqrt{1+\ell}\bigg[\omega&-\frac{mM\tilde{a}\sqrt{1+\ell}}{r_h^4}\bigg(2Mr_h\\
  &-\frac{e^{-\gamma}Q^2(2+\ell)}{2(1+\ell)}  \bigg)\bigg]+\mathcal{O}(\tilde{a}^2),
\end{aligned}
\end{equation}
and the parameter $q$ in the exponential term depends on the boundary condition applied at infinity, with $Re(q)>0$ for QNMs and $Re(q)<0$ for quasibound states.

Based on the two asymptotic solutions obtained above, we can construct a first-order slow-rotation ansatz that is consistent with the boundary conditions at both the event horizon and spatial infinity, namely,
\begin{align}\label{psia2}
\Psi_l(r)= & e^{-\sqrt{1+\ell}qr}(r-r_m)^{\sqrt{1+\ell}\chi}\left(\frac{r-r_h}{r-r_m}\right)^{-i\sqrt{1+\ell}\rho}R(r),
\end{align}
where 
\begin{align}
\chi&=\frac{M(2\omega^2-\mu^2)}{q},\\
\rho&=\frac{r_h^2}{r_h-r_m}\bigg[\omega-\frac{mM\tilde{a}\sqrt{1+\ell}}{r_h^4}\bigg(2Mr_h
  -\frac{Q^2(2+\ell)}{2e^{\gamma}(1+\ell)}  \bigg)\bigg].
\end{align}
In the subsequent subsections, we provide a detailed analysis of the QNMs of massive scalar perturbations by employing two complementary numerical techniques: first, the matrix method, which offers a flexible and efficient framework for handling a broad class of effective potentials, and second, the continued fraction method, which has proven to be highly accurate and widely adopted in black hole perturbation theory.

\subsection{Numerical methods}

In order to compute the QNM spectrum of massive scalar perturbations, we employ two complementary numerical techniques: the matrix method (MM) and the continued fraction method (CFM). Presenting them in this order highlights how the MM provides a flexible modern framework, while the CFM remains the benchmark for precision.

The MM was originally developed in a series of works by Lin and collaborators~\cite{Lin:2016sch,Lin:2017oag,Lin:2019mmf,Lin:2022ynv,Lei2021,Liu:2023uft}, and has been widely applied to perturbation problems in black hole physics. Its main advantage is that it does not rely on constructing a special trial series; instead, it only requires the enforcement of boundary conditions to yield accurate results.

We begin by introducing the compactified coordinate
\begin{align}\label{transyr}
y=\frac{r-r_h}{r-r_m}, 
\end{align}
which maps the radial domain to $y\in[0,1]$. To implement the correct boundary conditions, we redefine the wave function as
\begin{align}\label{transchipsi}
\chi(y)=y(1-y)\psi_l(y),
\end{align}
leading directly to 
\begin{align}\label{chi01}
\chi(0)=\chi(1)=0.
\end{align}
With this substitution, the perturbation equation takes the form
\begin{align}\label{qicieq}
\tilde{\mathcal{C}}_2(y,\omega)\chi''(y)+\tilde{\mathcal{C}}_1(y,\omega)\chi'(y)+\tilde{\mathcal{C}}_0(y,\omega)\chi(y)=0,
\end{align}
where the coefficients depend linearly on $\omega$, 
\[
\tilde{\mathcal{C}}_j(y,\omega)=\tilde{\mathcal{C}}_{j,0}(y)+\omega \tilde{\mathcal{C}}_{j,1}(y), \quad j=0,1,2.
\]
Discretizing the interval $y\in[0,1]$ into evenly spaced grid points and expanding $\chi(y)$ about each point, we obtain the differential matrices associated with Eq.~\eqref{qicieq}. The resulting algebraic system can be cast into the compact matrix form
\begin{align}
\left(\mathcal{M}_0+\omega\mathcal{M}_1\right)\chi(y)=0,
\end{align}
where $\mathcal{M}_0$ and $\mathcal{M}_1$ are $N\times N$ matrices determined by the discretization. 
Solving this eigenvalue problem provides the desired QNM spectrum.  

For comparison and validation, we also apply the CFM, introduced by Leaver~\cite{Leaver}, which is known for its remarkable accuracy in QNM calculations. In this method, the radial solution is expanded near the event horizon as a power series:
\begin{align}
R_l(r)=\sum_{n=0}^{\infty}d_n\left(\frac{r-r_h}{r-r_m}\right)^n .
\end{align}
Substituting this expansion into Eq.~\meq{eq111} leads to a seven-term recurrence relation for the coefficients. The first few terms can be written schematically as
\begin{equation}\label{grav0-4}
\begin{aligned}
d_1=&\,{\mathcal{C}}_{1,0}d_0,\\
d_2=&\,{\mathcal{C}}_{2,0}d_0+{\mathcal{C}}_{2,1}d_1,\\
d_3=&\,{\mathcal{C}}_{3,0}d_0+{\mathcal{C}}_{3,1}d_1+{\mathcal{C}}_{3,2}d_2,\\
d_4=&\,{\mathcal{C}}_{4,0}d_0+{\mathcal{C}}_{4,1}d_1+{\mathcal{C}}_{4,2}d_2+{\mathcal{C}}_{4,3}d_3,\\
d_5=&\,{\mathcal{C}}_{5,0}d_0+{\mathcal{C}}_{5,1}d_1+{\mathcal{C}}_{5,2}d_2+{\mathcal{C}}_{5,3}d_3+{\mathcal{C}}_{5,4}d_4,
\end{aligned}
\end{equation}
while the general recurrence relation takes the form
\begin{equation}\label{6oder}
\begin{aligned}
&d_{n+1}\alpha_n+d_{n}\beta_n+d_{n-1}\gamma_n+d_{n-2}\sigma_n \\
&\quad +d_{n-3}\tau_n+d_{n-4}\delta_n+d_{n-5}\epsilon_n=0,\qquad n=5,6,\dots
\end{aligned}
\end{equation}
with all coefficients depending on the parameters $l,m$, $\ell$, $\tilde{a}$, $M\mu$, $\gamma$, $Q/M$, $M\omega$ and $n$. 
Their explicit expressions are lengthy and omitted for brevity.  
By providing a sufficiently large value of $n$, we can solve for $\omega$ using these recurrence formulas.

In summary, the MM offers flexibility and ease of implementation, while the CFM provides benchmark accuracy. 
Using both methods in tandem allows us to cross-check results and ensure the robustness of the QNM spectrum obtained in the Lorentz-violating bumblebee background with conformal nonlinear electrodynamics.

\subsection{Numerical results}
\begin{figure}[h]
\centering
\includegraphics[width=1\linewidth]{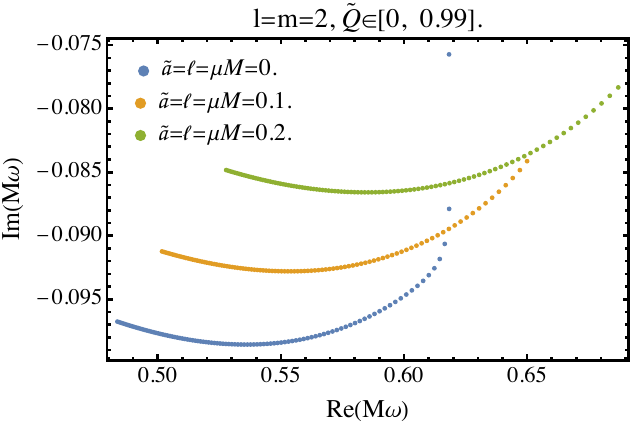}
\caption{Variation of the real and imaginary parts of the fundamental QNM frequency of the massive scalar field with $l=m=2$ as the parameter $ \tilde{Q} $  increases toward the extremal limit. }
\label{fig1a}
\end{figure}

In this subsection, we numerically calculate the QNMs using both the matrix method (of order 15) and the continued fraction method (of orders 10 and 20), and present the results in Fig. \ref{fig1a}. 
For simplicity, we set $M=1$ without loss of generality. 
We define an effective conformal nonlinear charge $\tilde{Q}=e^{-\gamma}Q^2/M^2$, which is a dimensionless quantity ranging from 0 to 1. 
In this paper, we focus on the fundamental modes with $l=m=2$, as they are among the most representative modes. 
For comparison with the study in Ref. \cite{Deng:2025uvp}, we present in Fig. \ref{fig1a} the trend of the QNM frequencies as the effective conformal nonlinear charge increases from 0 to near-extremal black hole values, for the case $\tilde{a}=\ell=\mu M=0/0.1/0.2 $.
Across varying spin, Lorentz-violation, and field-mass parameters, the influence of the effective conformal nonlinear charge on the QNM frequencies exhibits an almost identical trend: the real part of the frequency increases monotonically, while the imaginary part first decreases and then rises as the charge approaches its extremal value. 
And the parameter value step associated with near-extremal configurations becomes increasingly significant. 
Since the results obtained from the three numerical approaches are nearly indistinguishable, the data points representing the QNM frequencies in Fig. \ref{fig1a} completely overlap at the plotted scale. 
\begin{figure}[h]
\centering
\includegraphics[width=1\linewidth]{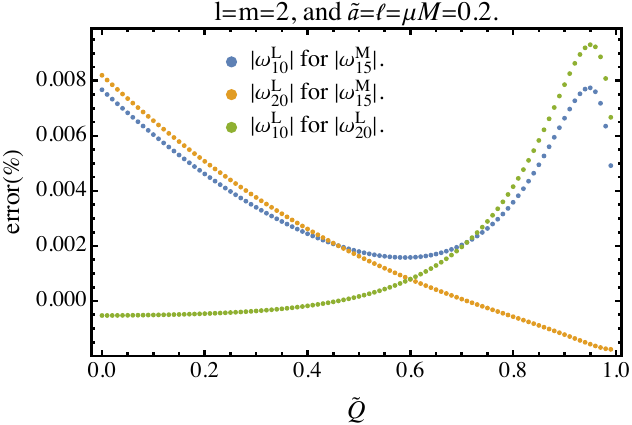}
\caption{
The percentage error between CFM (of orders 10 and 20) and MM (of order 15) results is analyzed. 
Data points are sampled at intervals of 0.01 for the dimensionless parameters $ \tilde{Q} $.}
\label{fig1b}
\end{figure}
To more clearly reveal the minor discrepancies among the methods, Fig. \ref{fig1b} presents the percentage errors in the QNM frequencies calculated by different approaches.
The error is defined as the relative difference between the magnitudes of the frequencies from two sets of data, A and B, as follows:
\begin{equation}
\text{error}=\frac{|\omega_A|-|\omega_B|}{|\omega_B|} \times 100,
\end{equation}
where A and B refer to the two different datasets used for comparison.
For brevity, we present only the case with $\tilde{a}=\ell=\mu M=0.2$. 
As shown in Fig. \ref{fig1b}, the errors for all parameter points are below $0.01\%$, which typically corresponds to an accuracy better than $10^{-4}$. 
Therefore, the results obtained from our numerical methods can be regarded as sufficiently precise.
Considering the balance between computational cost and numerical accuracy, all subsequent results are obtained using the continued fraction method at the 10th order ($n=10$).

\begin{figure}[h]
\centering
\includegraphics[width=1\linewidth]{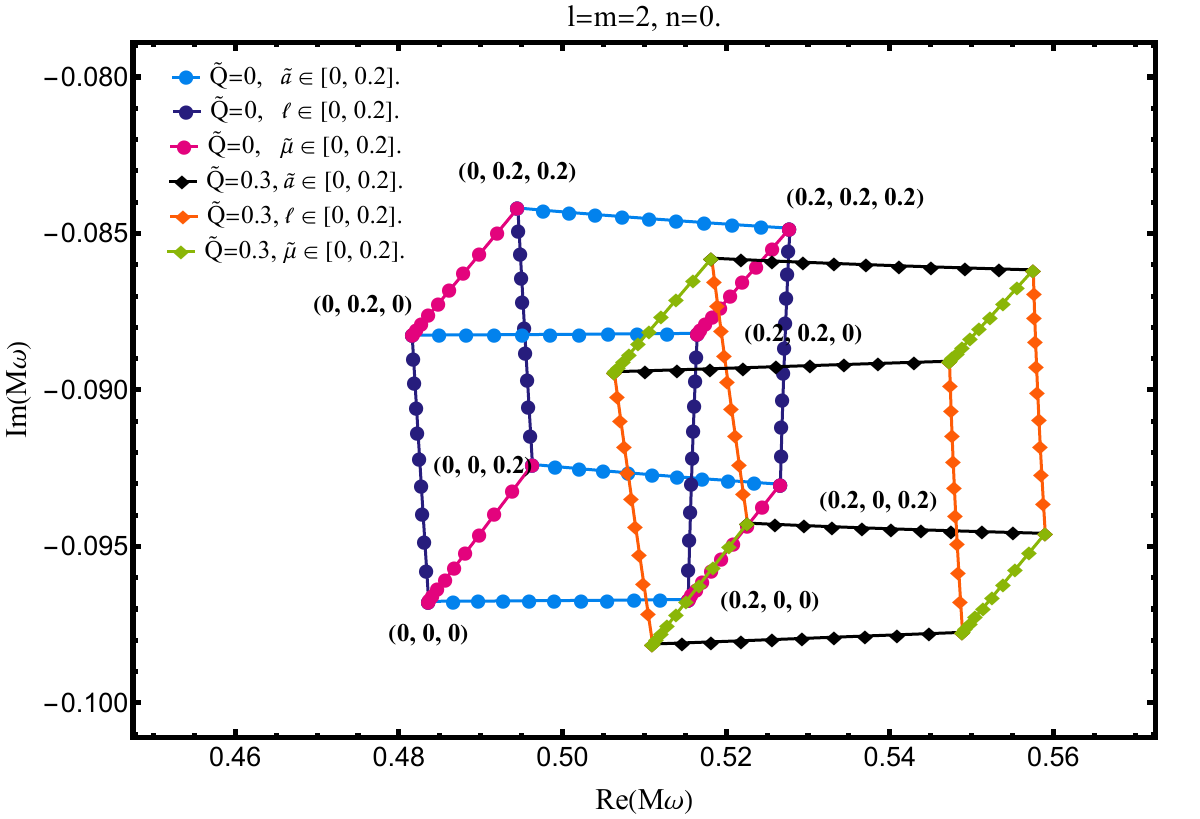}
\caption{We display the complex scalar frequencies of the $ l=m=2, n=0 $ modes as functions of the spin, Lorentz-violation parameter, and field mass, for two cases of the effective charge parameter: $ \tilde{Q}=0 $ (left) and $\tilde{Q}=0.3$ (right).
}
\label{fig2}
\end{figure}

\begin{figure}[h]
\centering
\includegraphics[width=1\linewidth]{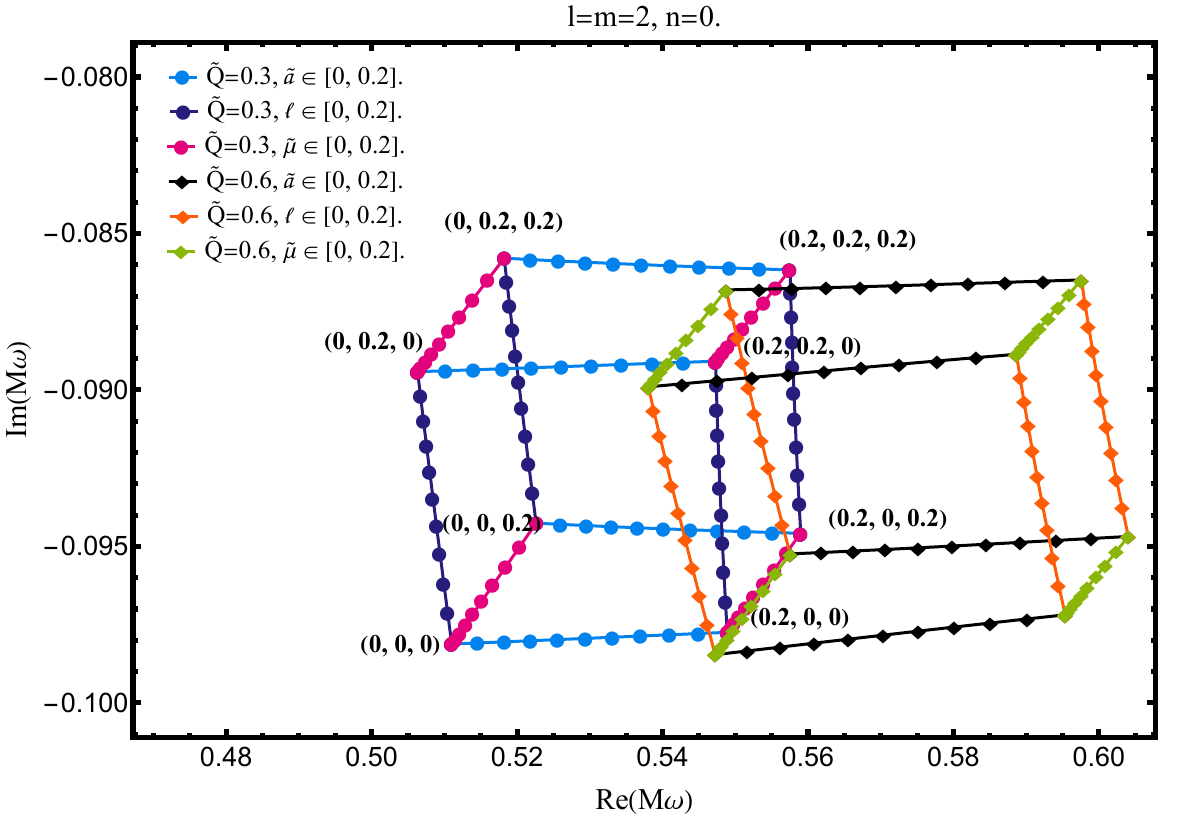}
\caption{We display the complex scalar frequencies of the $ l=m=2, n=0 $ modes as functions of the spin, Lorentz-violation parameter, and field mass, for two cases of the effective charge parameter: $ \tilde{Q}=0.3 $ (left) and $\tilde{Q}=0.6$ (right).}
\label{fig3}
\end{figure}

\begin{figure}[h]
\centering
\includegraphics[width=1\linewidth]{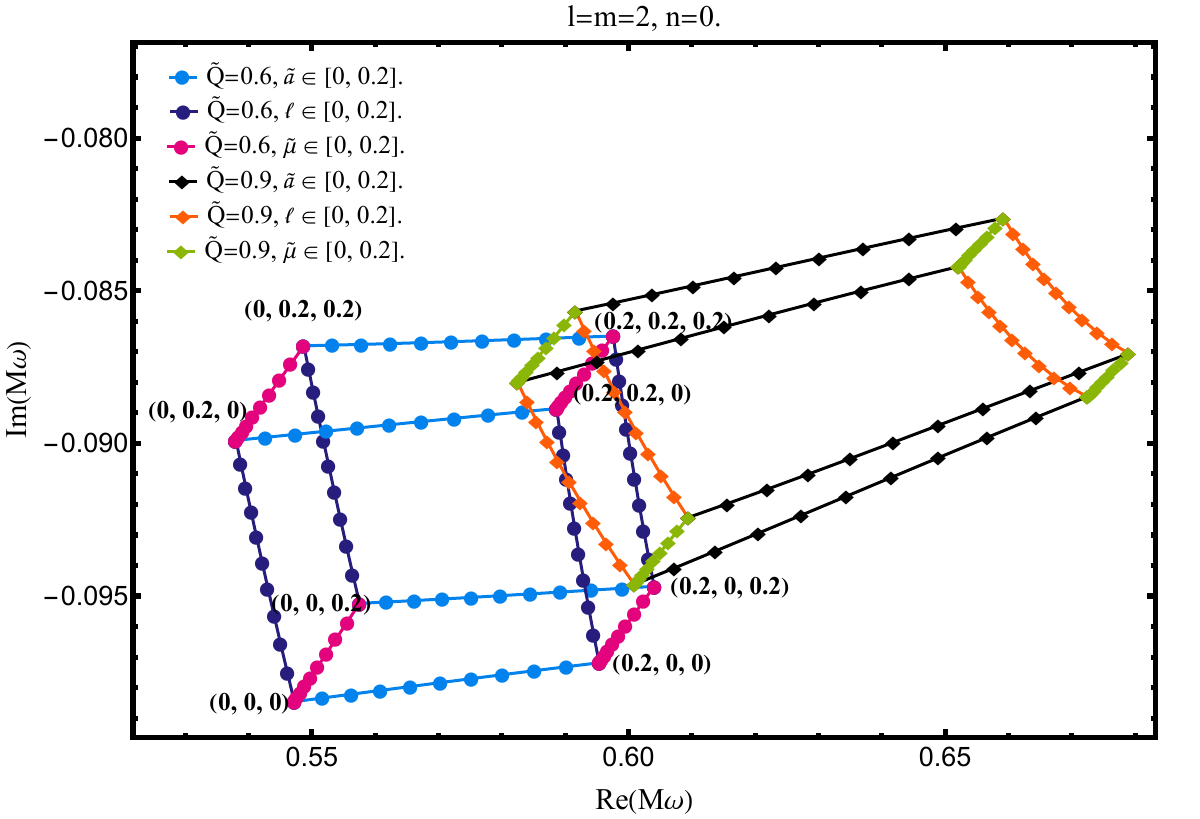}
\caption{We display the complex scalar frequencies of the $ l=m=2, n=0 $ modes as functions of the spin, Lorentz-violation parameter, and field mass, for two cases of the effective charge parameter: $ \tilde{Q}=0.6 $ (left) and $\tilde{Q}=0.9$ (right).}
\label{fig4}
\end{figure}

In the following, we perform a comprehensive analysis of how the four parameters $\tilde{a}$, $\ell$, $\tilde{\mu}$, and $\tilde{Q}$ affect the QNM frequencies.
In Figs. \ref{fig2}–\ref{fig4}, we display the complex QNM frequencies of the fundamental $l=m=2$ mode under simultaneous variations of the spin parameter $\tilde a$, the Lorentz-violation parameter $\ell$, and the field mass~$\mu M$ for several values of the effective conformal nonlinear charge $\tilde Q$. 
For each fixed $\tilde Q$, the frequency points corresponding to $(\tilde a,\ell,\mu M)=(0,0,0)\!\to\!(0.2,0.2,0.2)$ move coherently toward larger $\mathrm{Re}(M\omega)$ and smaller damping rates (i.e.\ smaller $|\mathrm{Im}(M\omega)|$). 
The combined variation of the three parameters produces the largest displacement of the complex frequencies, indicating that their impacts are approximately additive and co-directional within the examined parameter range.
As shown in Fig.\ref{fig2}, when $\tilde Q$ increases from 0 to 0.3, the entire cluster of frequency points migrates upward and to the right in the complex plane.  
This trend demonstrates that a stronger effective charge, or equivalently a weaker ModMax nonlinearity, enhances the oscillation frequency while slightly reducing the damping.  
All three parameters $(\tilde a,\ell,\mu M)$ exhibit monotonic, nearly linear influences on both the real and imaginary parts of the frequency, and no competing effects are observed.
Fig. \ref{fig3} extends this analysis to $\tilde Q=0.3$ and 0.6, revealing the same pattern: larger $\tilde Q$ systematically pushes the spectra toward higher $\omega_R$ and smaller $|\omega_I|$.  
The direction and magnitude of the frequency shifts caused by $\tilde a$, $\ell$, and $\mu M$ remain consistent, implying that the ModMax deformation and the Lorentz-violating background act in concert rather than in opposition.  
Physically, this trend originates from the effective charge term $\tilde{Q}=e^{-\gamma}Q^2/M^2 $ that governs both the metric and the potential.
An increase in $ \tilde{Q} $ effectively strengthens the Coulomb contribution, reshaping the near-horizon potential barrier and the phase structure of the scalar wave.
Finally, Fig. \ref{fig4} presents the cases $\tilde Q=0.6$ and 0.9.  
As the effective charge approaches its upper bound, the frequency loci continue to shift toward the upper-right region of the complex plane, though the relative displacement caused by further increases in $\tilde Q$ becomes progressively weaker, suggesting a mild saturation of the nonlinear contribution.  
Throughout all panels, the frequency evolution remains smooth and monotonic without any indication of mode branching or instability, confirming that the QNM spectrum retains a well-defined fundamental structure across the explored parameter domain.

Although the parameters $Q$ and $\gamma$ enter the spectrum only through the effective combination $\tilde Q = e^{-\gamma}Q^2/M^2$, the underlying field equations reveal that the ModMax nonlinearity governed by $\gamma$ introduces qualitatively distinct modifications to the spacetime geometry and the perturbation potential. 
As seen in Eqs. \meq{eq39} and \meq{eq69}, the exponential factor $e^{-\gamma}$ directly rescales the Coulomb term in the metric and the effective potential, thereby reducing the electromagnetic contribution to the curvature and lowering the height of the potential barrier. 
At the same time, the presence of both $e^{-\gamma}$ and $e^{\gamma}$ terms in the slow-rotation correction alters the phase structure of the propagating mode. 
Consequently, increasing $\gamma$ (stronger nonlinearity) weakens the near-horizon confinement and leads to larger damping rates, whereas decreasing $\gamma$ (weaker nonlinearity, larger $\tilde Q$) strengthens the oscillatory response and prolongs the lifetime of the perturbations.

\section{ CONCLUSIONS }\label{sec4}

In this work we have derived and analysed a novel family of electrically charged, slowly rotating black hole solutions in Einstein-bumblebee gravity minimally coupled to the traceless (conformal) ModMax nonlinear electrodynamics. 
By adopting a quadratic bumblebee potential we fix the vacuum expectation value of the Lorentz-violating vector and thereby promote spontaneous LSB to a controlled deformation of the gravitational sector. 
Working to first order in the rotation parameter we obtained both the static solutions and their slowly rotating generalisations, and we identified how the dimensionless bumblebee deformation $\ell$ and the ModMax parameter $\gamma$ enter the geometry and electromagnetic sector.

Technically, $\ell$ produces a uniform rescaling of the radial metric component that manifests as an asymptotically ``tilted'' vacuum and nontrivially alters causal structure and horizon multiplicity, while $\gamma$ appears in the electric sector through an exponential screening factor $e^{-\gamma}$ that continuously deforms the effective electric charge away from the Maxwell limit $(\gamma=0)$. 
The combined action of these two deformation parameters therefore controls both the spacetime geometry and the electromagnetic backreaction in a parametrically transparent way, providing a compact, four-parameter $(M,Q,\ell,\gamma)$ setting to probe their mutual interplay.
The QNM spectra exhibit a coherent and monotonic response to all physical parameters,  with larger effective charge or weaker nonlinearity leading to higher oscillation frequencies and smaller damping rates. 
These features highlight the potential observational relevance of Lorentz-violating and nonlinear electrodynamic effects in black hole ringdown signals.

Beyond the scope of the present analysis, several promising directions merit further investigation.
First, the spacetime constructed here offers a natural arena to examine quantum information theoretic aspects of Lorentz-violating and nonlinear electrodynamic backgrounds, such as the modification of entanglement harvesting or mutual information between Unruh–DeWitt detectors in curved spacetime \cite{Liu:2025zts,Wu:2025qqu,Tang:2025mtc,Liu:2025rks,Barros:2025xws,Wu:2025pxm,Li:2025bzd,Wu:2024whx,Wu:2024qhd,Wu:2022xwy,Du:2025ipb,Wang:2025lga,Liu:2023zro,Liu:2020jaj,Ji:2024fcq,Yu:2022bed}.
Second, it would be interesting to explore the thermodynamic and topological classification of these black holes, including how the spontaneous Lorentz-symmetry breaking and the ModMax nonlinearity jointly affect the phase structure, heat capacity, and topological charges associated with black hole entropy \cite{Wei:2022dzw,Wei:2024gfz,Wu:2024asq,Liu:2025iyl,Zhu:2024zcl,Wu:2024rmv,Wu:2023fcw,Wu:2023xpq,Wu:2023sue,Wu:2022whe,Ai:2025vno,Cunha:2024ajc}.
Finally, extending the present solution beyond the slow-rotation approximation or coupling it to additional matter sectors may provide valuable insights into astrophysical signatures and gravitational-wave ringdown templates within Lorentz-violating gravity theories.




\appendix

\begin{widetext}
\section{Energy-momentum tensor}\label{APPA}
In this appendix, we show the specific form of the energy-motion tensor in equation \meq{modified}, as follow
\begin{equation}
\begin{aligned}
T_{\mu\nu}^{\rm M} = &
\Biggr\{
\Bigl(-\cosh\gamma
+\frac{F_{\sigma\gamma}F^{\sigma\gamma}}
{\sqrt{(F_{\sigma\gamma}F^{\sigma\gamma})^{2}+(F_{\sigma\gamma}\tilde F^{\sigma\gamma})^{2}}}
     \sinh\gamma\Bigr) F_{\mu}{}^{\lambda}F_{\nu\lambda}
+\frac{F_{\sigma\gamma}\tilde F^{\sigma\gamma}\, 
\tilde F_{\mu}{}^{\lambda}F_{\nu\lambda}}
{\sqrt{(F_{\sigma\gamma}F^{\sigma\gamma})^{2}+(F_{\sigma\gamma}\tilde F^{\sigma\gamma})^{2}}}\sinh\gamma
      +\Bigl(-\tfrac14F_{\sigma\gamma}F^{\sigma\gamma}\cosh\gamma
\\
&\!+\!\tfrac14\sqrt{(F_{\sigma\gamma}F^{\sigma\gamma})^{2}+(F_{\sigma\gamma}\tilde F^{\sigma\gamma})^{2}}
     \sinh\gamma\Bigr)g_{\mu\nu}
\Biggl\}(1+\eta B_{\rho}B^{\rho})
\!+\!\frac{\eta}{2}
\bigg(
F_{\sigma\gamma}F^{\sigma\gamma}\,\cosh\gamma
\!-\!\sqrt{(F_{\sigma\gamma}F^{\sigma\gamma})^{2}+(F_{\sigma\gamma}\tilde F^{\sigma\gamma})^{2}}
     \sinh\gamma
\bigg)B_{\mu}B_{\nu}.  \label{Tm}
\end{aligned}
\end{equation}
and
\begin{equation}
\begin{aligned}
T^{\rm BB}_{\mu\nu} =& - \xi\biggr\{
-\tfrac12\,\nabla_{\alpha}\nabla_{\mu}(B^{\alpha}B_{\nu})
- \tfrac12\,\nabla_{\alpha}\nabla_{\nu}(B^{\alpha}B_{\mu})
+ \tfrac12\,\nabla^{2}(B_{\mu}B_{\nu})
+ \tfrac12\,g_{\mu\nu}\,\nabla_{\alpha}\nabla_{\beta}(B^{\alpha}B^{\beta})+B_{\mu}B^{\alpha}R_{\alpha\nu}\\
&
+B_{\nu}B^{\alpha}R_{\alpha\mu}
-\tfrac12\,g_{\mu\nu}\,B^{\alpha}B^{\beta}R_{\alpha\beta}
\biggl\}+2\,V'\,B_{\mu}B_{\nu}
- \Bigl(V + \tfrac{1}{4}B_{\sigma\gamma}B^{\sigma\gamma}\Bigr)g_{\mu\nu}
+ B_{\mu}{}^{\!\alpha}B_{\nu\alpha}.
\label{fa}
\end{aligned}
\end{equation}
where prime notation is adopted to indicate the derivative with respect to the argument of the relevant functions.
\end{widetext}

\section{Verification of the bumblebee equation \eqref{BB} to $\mathcal{O}(a)$.}
We verify Eq. \eqref{BB} by expanding each term in powers of $a$ and keeping only linear terms. The bumblebee equation can be written schematically as
\begin{equation}\label{Bumblebee_schematic}
\nabla_\mu B^{\mu\nu} - 2\Big(V'(B^2) B^\nu - \tfrac{\xi}{2}B^\mu R_\mu{}^\nu\Big) + \mathcal{T}^{\nu}_{\text{EM}\leftrightarrow B}=0,
\end{equation}
where $\mathcal{T}^{\nu}_{\text{EM}\leftrightarrow B}$ denotes the electromagnetic backreaction terms. Using the ansatz \eqref{eq:BumblebeeField} the background bumblebee vector is purely radial at $a=0$: $B_\mu=(0,B_r(r),0,0)$ + $\mathcal{O}(a^2)$. Important consequences follow immediately:

(i) The covariant divergence $\nabla_\mu B^{\mu\nu}$ for $\nu=t,\varphi$ contains factors proportional to $B^t,B^\varphi$ and/or derivatives of these components. Since $B^t=B^\varphi=0$ at the order considered, the resulting linear-in-$a$ elements for $\nu=t,\varphi$ vanish.

(ii) For $\nu=r$ the principal contribution to $\nabla_\mu B^{\mu r}$ is the radial derivative term present already in the static case,
\begin{equation}
\nabla_\mu B^{\mu r} = \frac{1}{\sqrt{-g}}\partial_r\big(\sqrt{-g}\,B^{r}\big) + \mathcal{O}(a^2),
\end{equation}
because the metric determinant $\sqrt{-g}$ and $B^{r}$ both change only at $\mathcal{O}(a^2)$ (recall $\rho^2=r^2+\mathcal{O}(a^2)$ and $\Delta_r=\Delta_0+\mathcal{O}(a^2)$). Hence the radial equation reduces to its static form up to $\mathcal{O}(a^2)$.

(iii) The curvature-coupling term $B^\mu R_\mu{}^\nu$ could in principle generate linear-in-$a$ components because off-diagonal Ricci components such as $R_{t\varphi}$ are $\mathcal{O}(a)$. However the contraction requires $B^\mu$ to have a time or azimuthal component in order to pick up these off-diagonal Ricci pieces. Since $B^t=B^\varphi=0$ to the working order, $B^\mu R_\mu{}^\nu=B^r R_r{}^\nu$ and $R_r{}^\nu$ has no linear-in-$a$ contribution that survives this contraction. Thus the curvature-coupling does not introduce a nonvanishing $\mathcal{O}(a)$ source.

(iv) Electromagnetic backreaction terms $\mathcal{T}^{\nu}_{\text{EM}\leftrightarrow B}$ involve products of $B_\mu$ (purely radial at this order) with electromagnetic invariants built from $F_{\mu\nu}$. Because the dominant electric field $F_{tr}$ is unchanged at $\mathcal{O}(a)$, these backreaction terms reproduce their static expressions at linear order and do not induce linear-in-$a$ violations.

Collecting (i)–(iv) yields that each component $\nu$ of Eq. \eqref{Bumblebee_schematic}\ vanishes at $\mathcal{O}(a)$; the first nonzero corrections are $\mathcal{O}(a^2\ell)$. Therefore Eq. \eqref{BB} is satisfied to linear order in the rotation parameter.

\subsection{Verification of the electromagnetic (ModMax) equation \eqref{FF} to $\mathcal{O}(a)$.}
Write Eq. (16) in divergence form
\begin{equation}\label{Maxwell_schematic}
\nabla_\nu\Big[\,\mathcal{C}(S,P,B^2)\,F^{\mu\nu} + \mathcal{D}(S,P,B^2)\,\tilde F^{\mu\nu}\,\Big] \;=\; 0,
\end{equation}
where $\mathcal{C}$ and $\mathcal{D}$ are the (nonlinear) scalars depending on the invariants $\mathcal{S}\equiv -\tfrac12 F_{\alpha\beta}F^{\alpha\beta}$ and $\mathcal{P}\equiv -\tfrac12 F_{\alpha\beta}\tilde F^{\alpha\beta}$, and also on the bumblebee invariant $B^2$ through the coupling $(1+\eta B^2)$. We perform an $\mathcal{O}(a)$ expansion.

1. \textbf{Expansion of field components.} From 
\begin{align}
&\begin{aligned}
A_t(r,\theta) = -\frac{e^{-\gamma}Q\,r}{\rho^2} = -\frac{e^{-\gamma}Q}{r} + \mathcal{O}(a^2), \label{At_expand}
\end{aligned}
\\
&\begin{aligned}
A_\varphi(r,\theta) &= a\,\frac{e^{-\gamma}Q\,\sqrt{1+\ell}\,r\sin^2\theta}{\rho^2}\\
&= a\,\frac{e^{-\gamma}Q\sqrt{1+\ell}\sin^2\theta}{r} + \mathcal{O}(a^3), \label{Aphi_expand}
\end{aligned}
\\
&\begin{aligned}
b_r(r,\theta) = b\,\frac{\rho}{\sqrt{1+\ell}\sqrt{\Delta_r}}
= b\,\frac{r}{\sqrt{1+\ell}\sqrt{\Delta_0(r)}} + \mathcal{O}(a^2).
\end{aligned}
\end{align}
we obtain
\begin{align}
F_{tr} &= \partial_r A_t = e^{-\gamma}Q\,\frac{1}{r^2} + \mathcal{O}(a^2), \label{Ftr}\\
F_{r\varphi} &= \partial_r A_\varphi = a\left(-e^{-\gamma}Q\sqrt{1+\ell}\,\frac{\sin^2\theta}{r^2}\right)+\mathcal{O}(a^3), \label{Frphi}\\
F_{t\varphi} &= \partial_\varphi A_t - \partial_t A_\varphi = \mathcal{O}(a^1)\ \text{(multipole structure)}.
\end{align}
Thus the principal electric component $F_{tr}$ is unchanged at $\mathcal{O}(a)$ while the rotation induces a magnetic-type component $F_{r\varphi}\propto a$.

2. \textbf{Invariants $\mathcal{S}$ and $\mathcal{P}$.} Because $\mathcal{S}$ at zeroth order is built from $F_{tr}^2$ and the leading $a$-correction to $F_{tr}$ is $\mathcal{O}(a^2)$, we have
\begin{equation}
\mathcal{S} = \mathcal{S}_0(r) + \mathcal{O}(a^2), \qquad \mathcal{P} = \mathcal{O}(a),
\end{equation}
but $\mathcal{P}$ is an axial pseudoscalar typically proportional to $F\wedge F$ and in the present ansatz it is generated by the product $F_{tr}F_{r\varphi}$ which is proportional to $a\cdot F_{tr}^2/r^2$. However, $\mathcal{D}(S,P,B^2)$ multiplies $\tilde F^{\mu\nu}$ and its contribution to the $\mu=t$ sector is suppressed by angular integrals / structures that vanish identically for the present axisymmetric, purely radial background; furthermore $\mathcal{D}$ itself evaluates to $\mathcal{D}_0 + \mathcal{O}(a^2)$. Therefore, both $\mathcal{C}$ and $\mathcal{D}$ are effectively independent of $a$ in linear order:
\begin{equation}
\mathcal{C} = \mathcal{C}_0(r) + \mathcal{O}(a^2), \qquad \mathcal{D} = \mathcal{D}_0(r) + \mathcal{O}(a^2).
\end{equation}

3. \textbf{Check $\mu=t$ component.} The $\mu=t$ Eq. \eqref{Maxwell_schematic} reduces in leading order to
\begin{equation}
\frac{1}{\sqrt{-g}}\partial_r\big(\sqrt{-g}\,\mathcal{C}_0\,F^{tr}\big) + \mathcal{O}(a^2) = 0,
\end{equation}
because mixed components that could carry $\mathcal{O}(a)$ contributions are either multiplied by $B^\varphi,B^t$ (zero in this order) or are total angular derivatives that vanish upon using axisymmetry. Since $F^{tr}$ is unchanged at $\mathcal{O}(a)$ and $\sqrt{-g}=\sqrt{-g}_0+\mathcal{O}(a^2)$, the $\mu=t$ equation is satisfied to linear order if and only if the static radial equation holds; but the static radial equation is exactly the one used to fix the nonrotating solution. Hence the $\mu=t$ Maxwell/ModMax equation is satisfied to $\mathcal{O}(a)$.

4. \textbf{Check $\mu=\varphi$ component.} The $\varphi$-component contains the rotation-induced magnetic term:
\begin{equation}
\nabla_\nu\big(\mathcal{C}_0 F^{\varphi\nu}\big) \simeq \frac{1}{\sqrt{-g}_0}\partial_r\big(\sqrt{-g}_0\,\mathcal{C}_0\,F^{\varphi r}\big) + \mathcal{O}(a^2).
\end{equation}
Using the leading-order inverse-metric scalings $F^{\varphi r}\sim g^{\varphi\varphi}g^{rr}F_{r\varphi}\sim F_{r\varphi}/(r^4\sin^2\theta)$ and the explicit form \eqref{Frphi} for $F_{r\varphi}\propto a/r^2$, the combination $\partial_r(\sqrt{-g}_0\,\mathcal{C}_0\,F^{\varphi r})$ organizes into the total radial derivative that vanishes because we have chosen $A_\varphi$ with the Kerr–Newman functional dependence $A_\varphi\propto a\,r/\rho^2$. Concretely, the radial derivative acts on the factor $r^{-2}$ in $F_{r\varphi}$ and is canceled by the radial derivative of $\sqrt{-g}_0\,g^{\varphi\varphi}g^{rr}$ (the same cancellation occurs in the linearized Kerr–Newman-Maxwell solution). Therefore, the $\varphi$-equation holds at $\mathcal{O}(a)$.

Consequently, combining the analyzes $\mu=t$ and $\mu=\varphi$  and using axisymmetry, we conclude that Eq. \eqref{FF} is satisfied to linear order in $a$; the first nonvanishing deviations appear only at $\mathcal{O}(a^2\ell)$. Thus, the chosen $A_\varphi$ precisely supplies the rotation-induced magnetic component required to cancel the linear-in-$a$ terms of the divergence in Eq. \eqref{Maxwell_schematic}.

\bibliography{Reff}
\end{document}